\documentclass{PoS}

\title{GLAST Status and Application to Microquasars}

\ShortTitle{GLAST Status and Application to Microquasars}

\author{\speaker{Richard Dubois}%
         \thanks{This work is supported by Stanford University and the Stanford Linear Accelerator Center (SLAC) under DoE contract number DE-AC03-76SFO0515. }\\
        Stanford Linear Accelerator Center, Menlo Park CA, USA\\
        E-mail: \email{richard@slac.stanford.edu}}

\author{for the LAT Collaboration}
\abstract{The Gamma-ray Large Area Space Telescope (GLAST) is a next generation high energy gamma-ray observatory due for launch in Fall 2007. The primary instrument is the Large Area Telescope (LAT), which will measure gamma-ray flux and spectra from 20 MeV to > 300 GeV and is a successor to the highly successful EGRET experiment on CGRO. The LAT will have better angular resolution, greater effective area, wider field of view and broader energy coverage than any previous experiment in this energy range. An overview of the LAT instrument design and construction is presented which includes performance estimates with particular emphasis on how these apply to strudies of microquasars.  The nature and quality of the data that will be provided by the LAT is described with results from recent detailed simulations that illustrate the potential of the LAT to observe gamma ray variability and spectra.}

\FullConference{VI Microquasar Workshop: Microquasars and Beyond\\
		 September 18-22, 2006\\
		 Como, Italy}

\begin{document}

\section{Introduction}

GLAST is a next generation high-energy gamma-ray observatory designed
for making observations of celestial gamma ray sources in the energy
band extending from 20 MeV to more than 300 GeV. It follows in the
footsteps of the Compton Gamma Ray Observatory EGRET experiment, which
was operational between 1991-1999. The GLAST Mission is part of NASA's 
Office of Space and Science Strategic Plan, with launch anticipated in 
2007. The principal instrument of the GLAST mission is the Large Area 
Telescope (LAT) that is being developed jointly by NASA and the US
Dept. of Energy (DOE) and is supported by an international
collaboration of 26 institutions lead by Stanford University. 

\begin{figure}[htb]
\includegraphics[width=120mm]{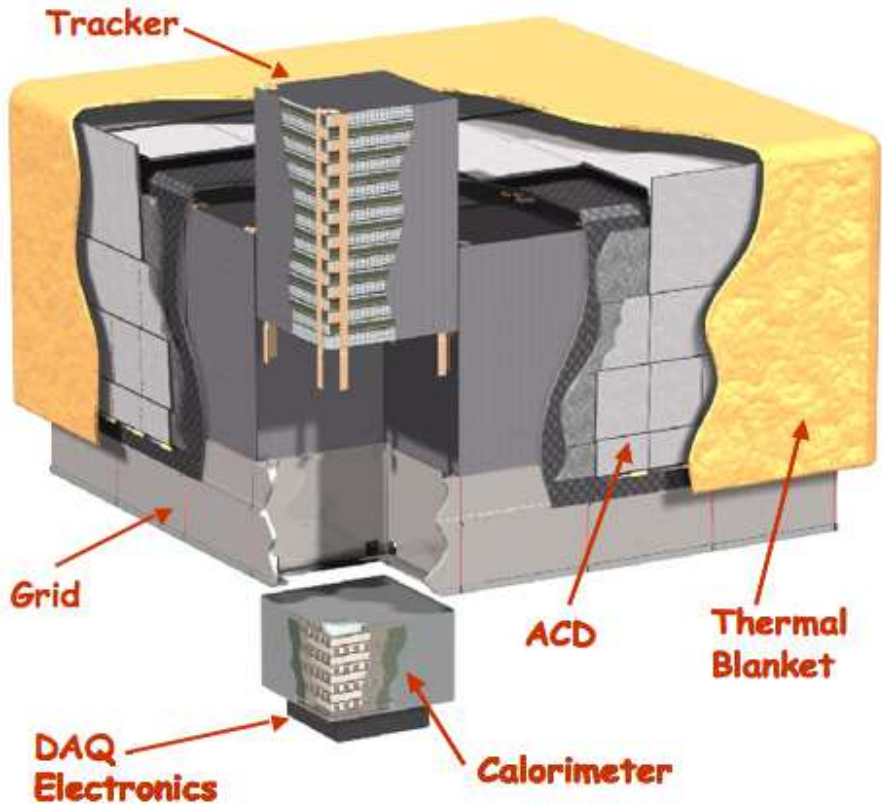}
\caption{Large Area Telescope. Composed of 16 modules containing
Tracker and Calorimeter elements, all surrounded by a scintillator
anti-coincidence shield.}
\label{LAT}
\end{figure}

The GLAST LAT is a high-energy pair conversion telescope that has been
under development for over 10 years with support from NASA, DOE and 
international partners. It consists of a precision converter-tracker, 
CsI hodoscopic calorimeter, plastic scintillator anticoincidence
system and a data acquisition system. The design is modular with a 4x4
array of identical tracker and calorimeter modules. The modules are 
approximately 38 x 38 cm. Figure \ref{LAT} shows the LAT instrument concept.

The LAT science instrument consists of an Anti Coincidence Detector 

(ACD), a silicon-strip detector Tracker (TKR), a hodoscopic CsI 
Calorimeter (CAL), and a Trigger and Data Flow system (T\&DF). The 
principal purpose of the LAT is to measure the incidence direction, 
energy and time of cosmic gamma rays while rejecting background from 
charged cosmic rays and atmospheric albedo gamma rays and particles. 
The data, filtered by onboard software triggers, are streamed to 
the spacecraft for data storage and subsequent transmittal to
ground-based analysis centers. The Tracker provides the principal
trigger for the LAT, converts the gamma rays into electron-positron 
pairs, and measures the direction of the incident gamma ray from 
the charged-particle tracks. The first 12 layers of converter are about 3\% X$_0$, while the next 4 layers (back) are about 18\%. This was done to optimize efficiency for photons interacting near the calorimeter.  The tracker is crucial in the first levels of
 background rejection for providing track information to extrapolate
 cosmic-ray tracks to the ACD scintillator tiles, and it is important 
for further levels of background analysis due to its capability 
to provide highly detailed track patterns in each event.

In normal operations the LAT will continually scan the sky, obtaining essentially complete sky coverage every 3 hours (two orbits).  This uniformity of sky coverage together with the large effective area and good angular resolution should permit many advances in the study of microquasars in the GeV range. The performance properties of the LAT are summarized in Table \ref{resp}. The most current LAT performance specifications are kept online\cite{LATPerf}.

\begin{table}[htdp]
\caption{LAT Energy and Angular Resolution}
\centering
\begin{tabular}{cc}Energy Resolution & $\approx$10\% ($\approx$5\% off axis) \\PSF (68\%) at 100 MeV & 3.5$^0$ front; 5$^0$ total \\PSF (68\%) at 10 GeV & 0.1$^0$ \\Field of View & 2.4 sr \\Point Source Sensitivity (> 100 MeV) & 3x10$^{-9}$ cm$^{-2}$ s$^{-1}$\end{tabular}
\label{resp}
\end{table}%

\section{Feasibility Study for Microquasar Observations}

GLAST's Data Challenge 2 (DC2) provided a detailed simulation of the sky and the LAT's response. 5 X-ray binaries were included with flux and spectra from EGRET measurements and known orbital periods. Fifty-five days of simulated orbit were created using a  full GEANT4 simulation of LAT response, with full time dependence in the simulations for AGN, solar flares, GRBs. It took some 200k CPU hours to produce this rendition of the sky. Modeling of the XRB's was very simple, with single power law energy spectra and full modulation of the orbital dependence.

For initial studies into GLAST's capabilities in observing microquasars we have concentrated on the two which have been measured in TeV gamma rays, by HESS\cite{HESS} and MAGIC\cite{MAGIC}. LS 5039 is near the galactic center and has two other nearby sources as well as a significant galactic diffuse background. On the other hand, LSI +61 303 is more isolated and in a lower diffuse background. In addition, their orbital periods differ by almost an order of magnitude. These provide good examples of the extremes of observations the LAT will be called on to make. We will investigate the LAT's ability to determine the flux, spectral index and orbital time variability in this pilot study.

\subsection{Full simulation 55 day orbit data}

\begin{figure}[hbtp]
\centering
\includegraphics[height=40mm]{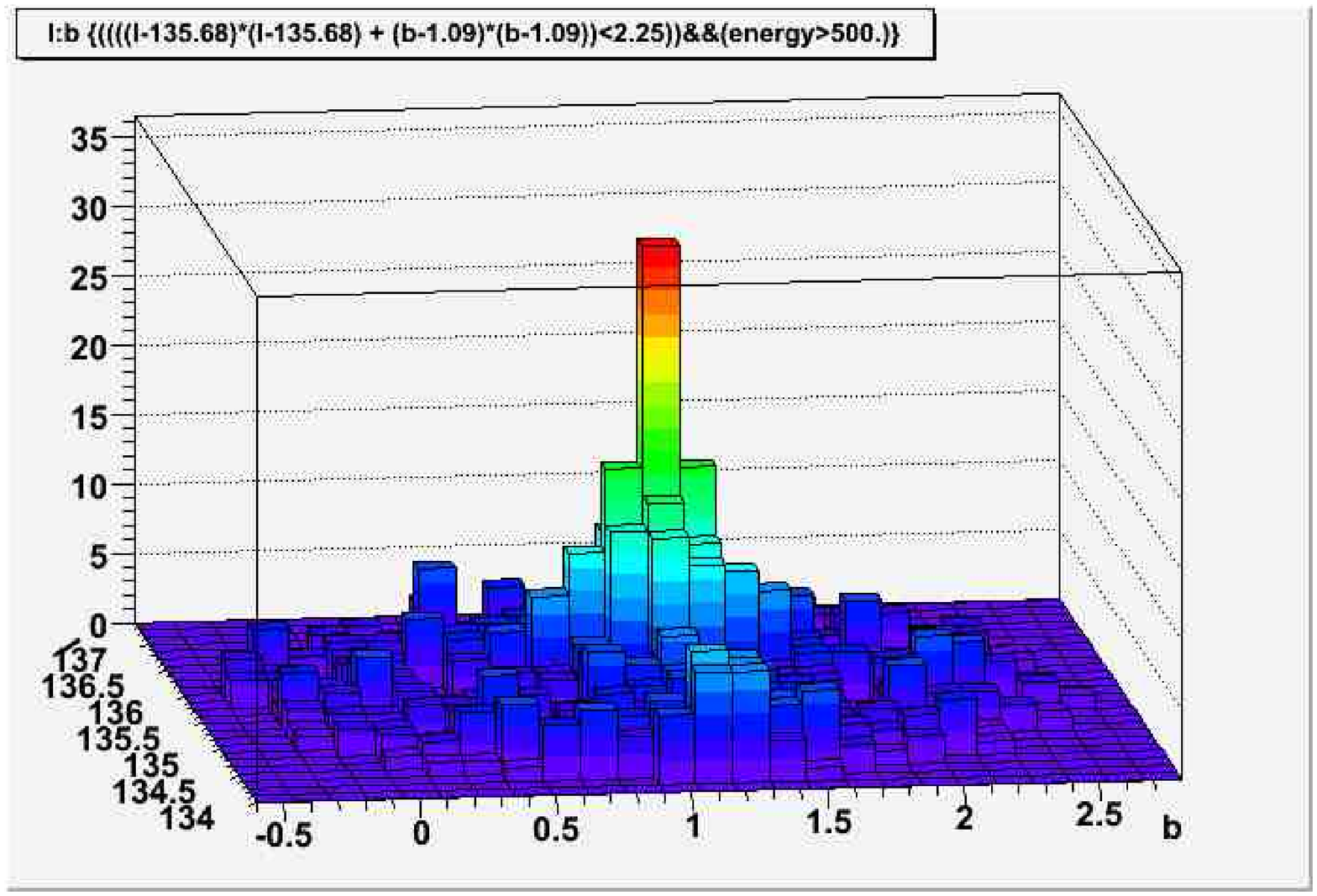}
\includegraphics[height=40mm]{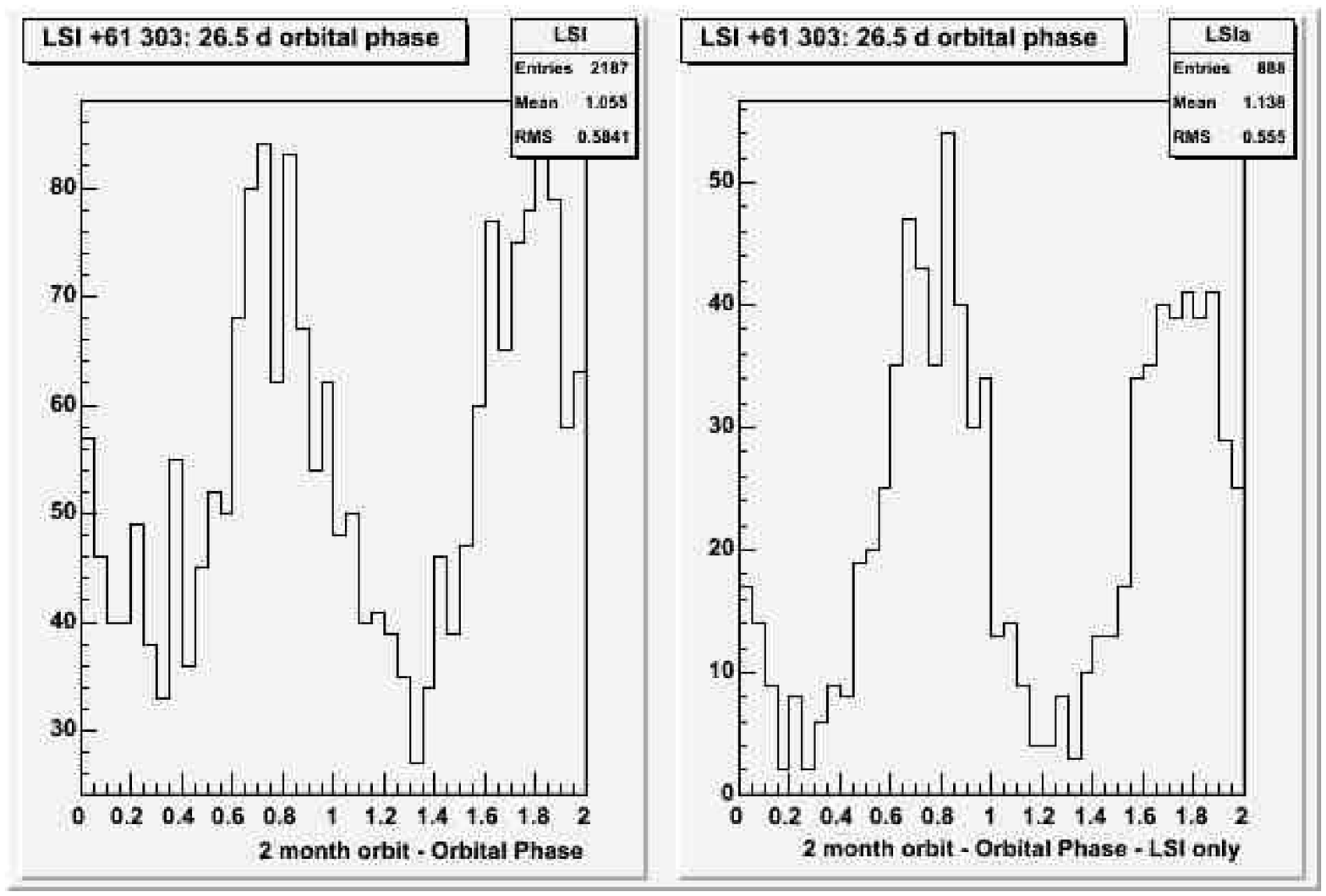}
\caption{Left: Galactic coordinates $l$ vs $b$ for LSI +61 303 for 55 days' orbit data. No other known sources are within a few degrees, and galactic diffuse background is low. LSI +61 303 orbital phase for E > 100 MeV. Middle panel shows all photons. Right panel shows photons from LSI +61 303 only.}
\label{LSIlb}
\end{figure}

LSI +61 303 is shown to be isolated and in a low background region as shown in Figure \ref{LSIlb}. It is straightforward to use a likelihood fit on this region to include the galactic diffuse background as well as a point source with a single power law spectrum. Such a fit results in a spectral index of 2.21$\pm$ 0.07, compared to the simulated value of 2.19. A fit to the l vs b distribution yields a one-sigma error of about 0.02$^0$ in each variable.

Figure \ref{LSIlb} also shows the orbital period phase for two periods for both all photons around LSI +61 303, and the from the source alone. The full modulation is clearly seen, and also is plainly visible for the all photon case, demonstrating the ability to detect the 29.5 day period. 

\begin{figure}[htbp]
\centering
\includegraphics[height=40mm]{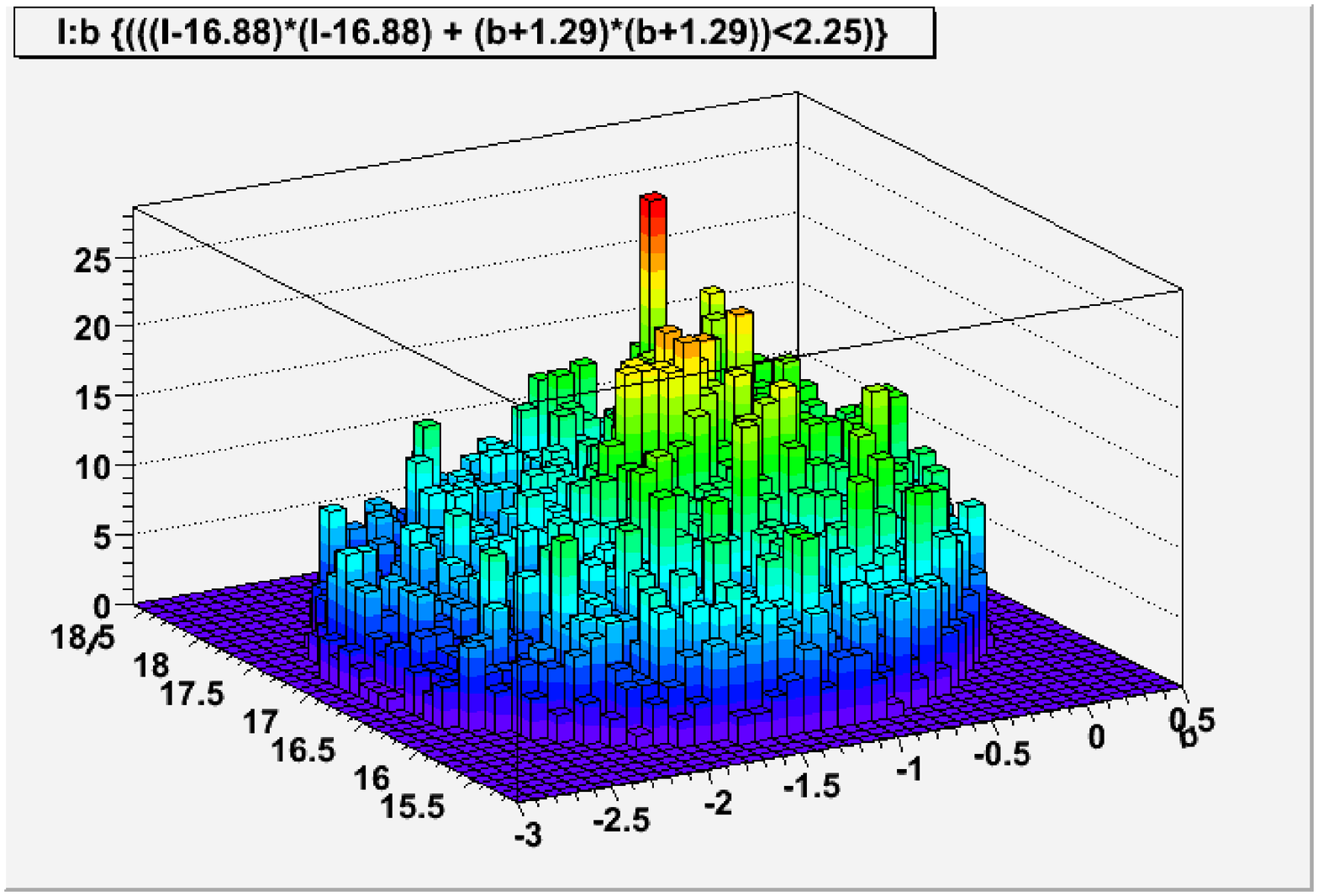}
\includegraphics[height=40mm]{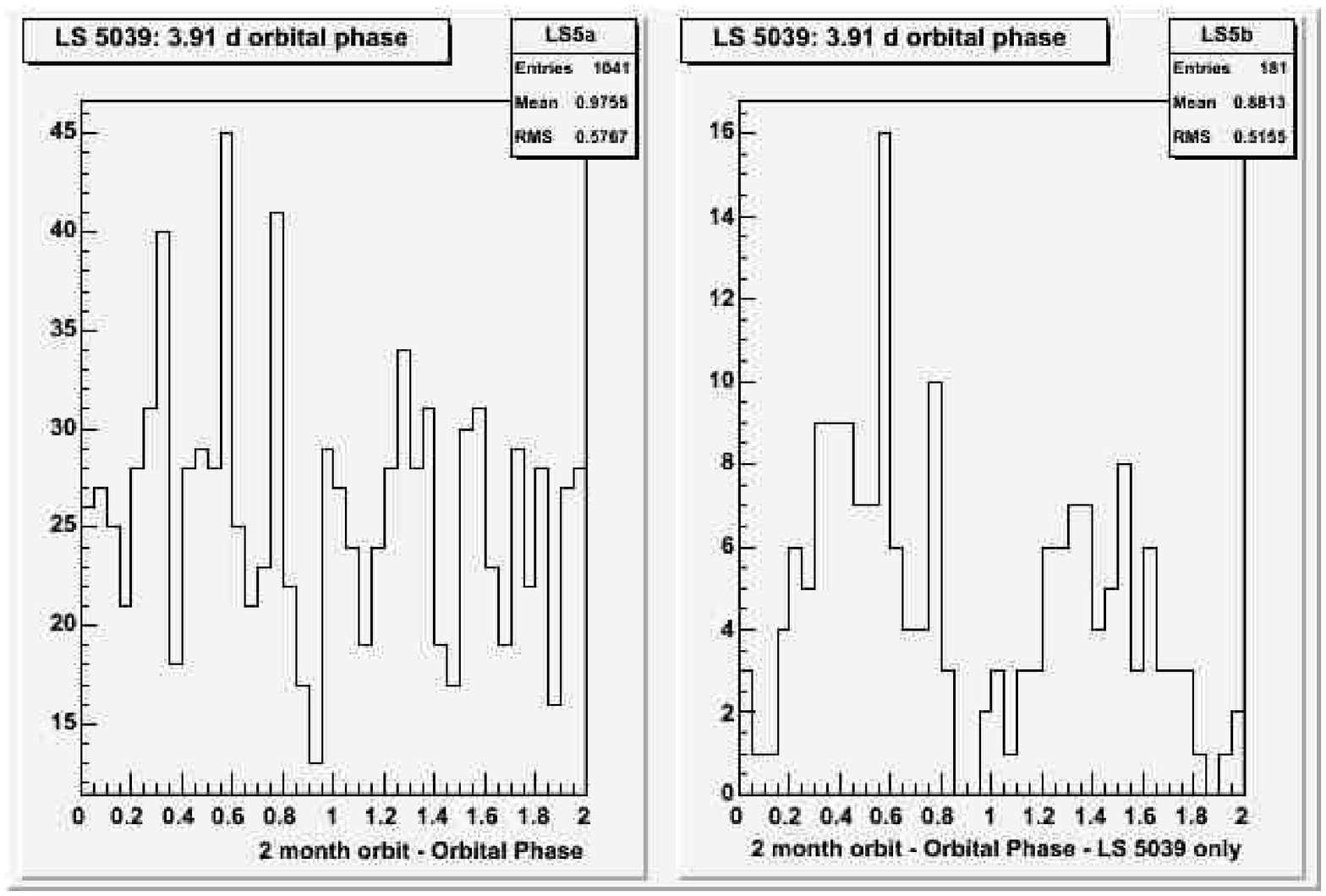}
\caption{Left: Galactic coordinates $l$ vs $b$ for LS 5039 for 55 days' orbit data.  There is no sign of LS 5039 at the generated source location (16.9$^0$,-1.3$^0$). Orbital phase for E > 100 MeV. Middle panel shows all photons. Right panel shows photons from LS 5039 only.}
\label{LS5lb}
\end{figure}

 Things are not so rosy for LS 5039, as shown in Figure \ref{LS5lb}. The background to noise ratio is about 15:1, so extracting the source photons will be very difficult. An attempt was made to use the PERIOD\cite{PERIOD} package, utilizing Lomb-Scargle periodigrams, to find the orbital period: for the LS 5039 photons alone, the orbital period was found to within 5\%; for the full photon sample, the process did not converge. 
 
\subsection{Fast simulation one year orbit data}

\begin{figure}[htbp]
\centering
\includegraphics[height=40mm]{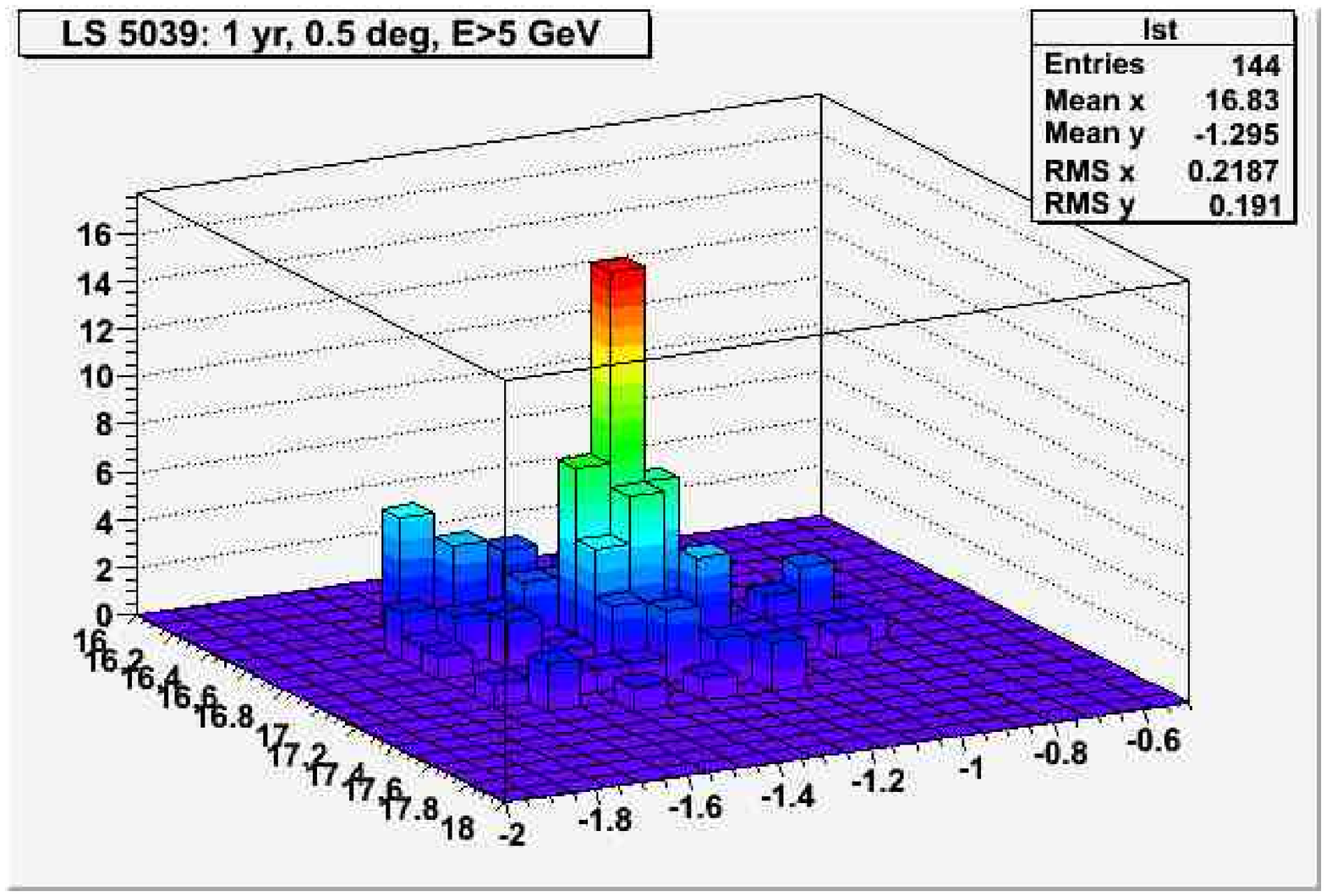}
\includegraphics[height=40mm]{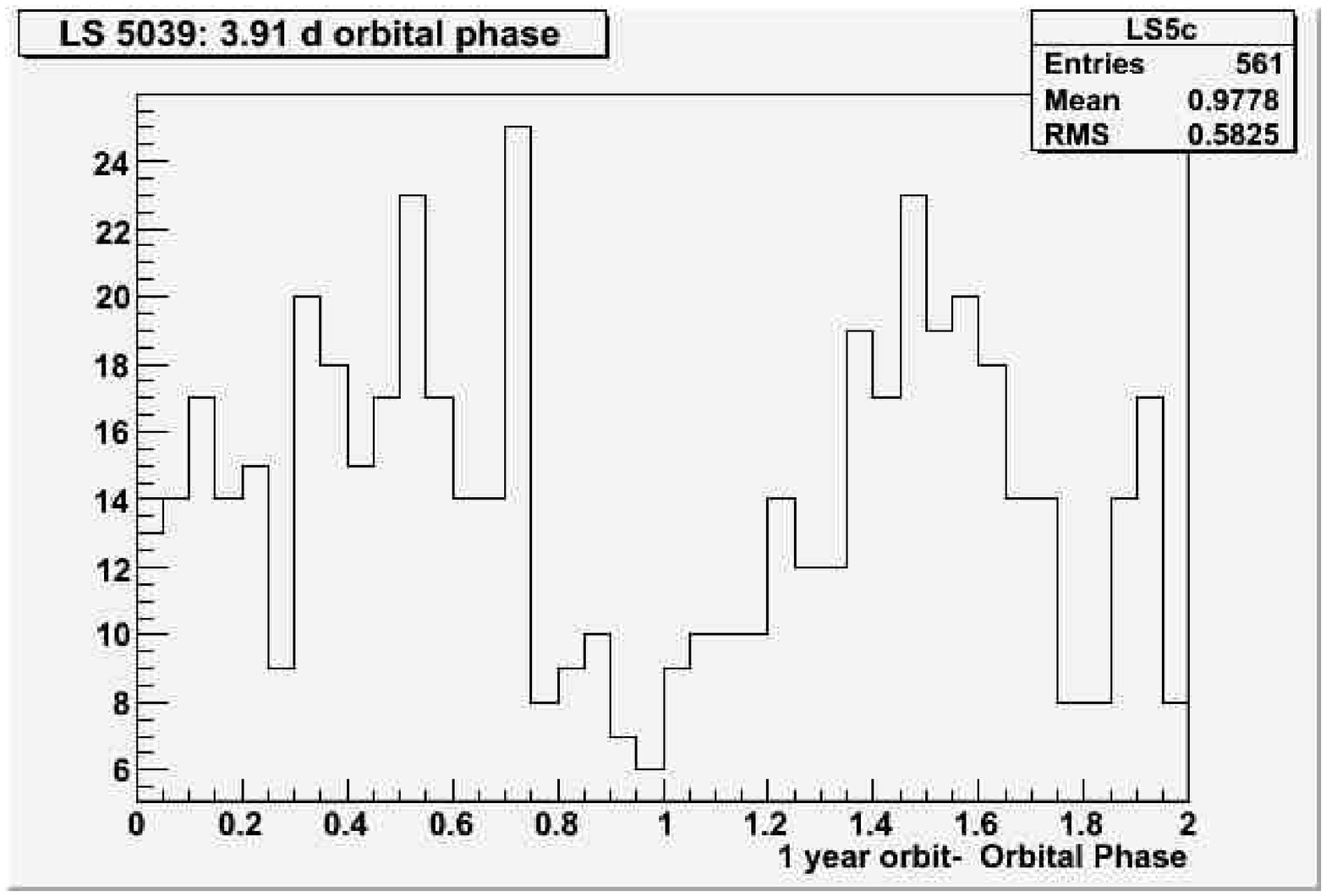}
\caption{Left: Galactic coordinates $l$ vs $b$ for LS 5039 for one year's' orbit data, E > 2 GeV.  LS 5039 is now quite evident. Orbital phase for E > 2 GeV. MIddle panel shows all photons. Right panel shows photons from LS 5039 only.}
\label{LS5year}
\end{figure}

 A full year of orbit data was simulation using a parametrization of the instrument response and effective area in order to investigate how much more data improves the situation for LS 5039: we will take advantage of the spectrum being harder than that of the diffuse and that at higher energies the photon directions are much better determined. This is illustrated in Figure \ref{LS5year}. In this case, the source is easily found and the 3.91 day orbital period has become apparent. A fit to the l vs b distribution yields an error of about 0.05$^0$ in each variable.
 
\section{Conclusions}

GLAST will launch in late 2007 and provide a unique view of the gamma ray sky over the 0.1-200 GeV energy range. The large field of view and survey mode of the LAT will allow us to monitor all the microquasar candidates on a continuous basis.

For isolated sources, we will have no difficulty measuring the spectral and orbital period properties for sources with fluxes like those of LSI +61 303 and LS 5039 ($\approx$10$^{-7}$ ph s$^{-1}$cm$^{-2}$). For sources living in confused regions of the sky, like LS 5039, we will need to integrate longer and make harder energy cuts to bring the signal out above background. A better source localisation could in principle be achieved with a likelihood analysis that took into account that the PSF is a strong function of photon energy.

Prior to launch we plan to extend our full simulations to a full year of orbit data and develop the machinery to observe all the microquasar candidates. We will also investigate our ability to detect time variabilities other than orbital and develop strategies for multi-wavelength campaigns.

\end{document}